\documentclass[%
 reprint,
 superscriptaddress,
 amsmath,
 amssymb,
prl,
]{revtex4-1}

\usepackage{graphicx}
\usepackage{dcolumn}
\usepackage{bm}

\usepackage{fullpage}
\usepackage{color}
\usepackage{subcaption}
\usepackage{multirow}
\usepackage{amsmath}
\usepackage{amssymb}
\usepackage{amsthm}
\usepackage[colorlinks   = true, urlcolor     = cyan, linkcolor    = blue, citecolor   = red]{hyperref}
\usepackage{bm}
\usepackage{pict2e}
\usepackage{epstopdf}
\usepackage{comment}
\usepackage{epsfig}
\usepackage{url}
\usepackage[ruled,vlined]{algorithm2e}

\newcommand{\beq}{\begin{equation}}
\newcommand{\eeq}{\end{equation}}
\newcommand{\beqs}{\begin{eqnarray}}
\newcommand{\eeqs}{\end{eqnarray}}
\newcommand{\beql}{\begin{equation} \label}

\usepackage{cancel}
\usepackage{amsmath}
\usepackage{amssymb}
\usepackage{etoolbox}
\usepackage{graphicx}
 \usepackage[english]{babel}
\usepackage{mathrsfs} 
\usepackage{enumitem}                                  

\usepackage{amsthm} 
\usepackage{siunitx}
\usepackage{ upgreek }

\newcommand{\toVect}[1]{\boldsymbol{#1}} 
\newcommand{\toMat}[1]{\mathbf{#1}}      

\DeclareMathAlphabet{\mathdutchcal}{U}{dutchcal}{m}{n}
\newcommand{\radialpolarization}{{\mathdutchcal{p}}}
\newcommand{\x}{{\toVect{\mathrm{x}}}}
\newcommand{\X}{{\toVect{\mathrm{X}}}}
\newcommand{\E}{{\toVect{\mathrm{E}}}}
\newcommand{\F}{{\toMat{F}}}
\renewcommand{\gg}{{G}}

\newcommand{\GG}{{\toMat{\gg}}}

\newcommand{\p}{{\toVect{\mathrm{p}}}}

\newcommand{\dd}{\text{\,}\mathrm{d}}

\newcommand{\n}{\toVect{\mathrm{n}}}

\newcommand{\Flexo}{\toVect{\flexo}}
\newcommand{\flexo}{\mu}

\newcommand{\norm   }[1]{\lVert #1 \rVert}

\begin{document}


\title{Transversal flexoelectric coefficient for nanostructures at finite deformations from first principles}

\author{David Codony}
\affiliation{Laboratori de C\`{a}lcul Num\`{e}ric, Universitat Polit\`{e}cnica de Catalunya, Barcelona, E-08034, Spain}

\author{Irene Arias}
\affiliation{Laboratori de C\`{a}lcul Num\`{e}ric, Universitat Polit\`{e}cnica de Catalunya, Barcelona, E-08034, Spain}
\affiliation{Centre Internacional de Metodes Numèrics en Enginyeria (CIMNE), 08034 Barcelona, Spain}

\author{Phanish Suryanarayana}
\email{phanish.suryanarayana@ce.gatech.edu}
\affiliation{College of Engineering, Georgia Institute of Technology, Atlanta, GA 30332, USA}

\date{\today}

\begin{abstract}
We present a novel formulation for calculating the transversal flexoelectric coefficient of nanostructures at finite deformations from first principles. Specifically, we introduce the concept of \emph{radial polarization} to make the coefficient a well-defined quantity for uniform bending deformations. We use the framework to calculate the flexoelectric coefficient for group IV atomic monolayers using density functional theory. We find that graphene's coefficient is significantly larger than previously reported, with a charge transfer mechanism that differs from other members of its group.  
\end{abstract}


\maketitle

\emph{Introduction.} Flexoelectricity \cite{tagantsev1991electric, yudin2013fundamentals, zubko2013flexoelectric, nguyen2013nanoscale, ahmadpoor2015flexoelectricity, krichen2016flexoelectricity, wang2019flexoelectricity}  is an electromechanical property common to insulating systems that represents a two-way coupling between strain gradients and polarization. In contrast to piezoelectricity, it is not restricted to materials with a specific symmetry, and in contrast to electrostriction, it permits reversal of the strain by reversal of the electric field. Due to the possibility of large strain gradients, the flexoelectric effect is particularly significant in nanostructures, making them ideal candidates for a number of  applications, including energy harvesting, sensing and actuating. 

A fundamental obstacle in characterizing and exploiting the flexoelectric effect is the significant disagreement between theory and experiment, with coefficients differing by up to three orders of magnitude, and sometimes even in the sign \cite{zubko2013flexoelectric, hong2010flexoelectricity,yudin2013fundamentals}. In view of this, perturbative approaches in the framework of Kohn-Sham density functional theory (DFT) \cite{KohnSham_DFT} have been developed  for calculating the flexoelectric tensor components  from first principles \cite{hong2011first, hong2013first, stengel2013flexoelectricity, stengel2014surface, dreyer2018current}. However, the coefficients so computed, of which the transversal component $\mu_\text{T}$ is particularly important for nanostructures, correspond to the asymptotic zero strain gradient limit. Therefore, they are restricted to linear response, likely not representative at the relatively large curvatures commonly encountered in experimental investigations involving bending deformations \cite{lindahl2012determination,chen2015bending, qu2019bending, han2019ultrasoft, han2019ultrasoft}.

\begin{figure}[htbp!]
\includegraphics[width=0.9\linewidth]{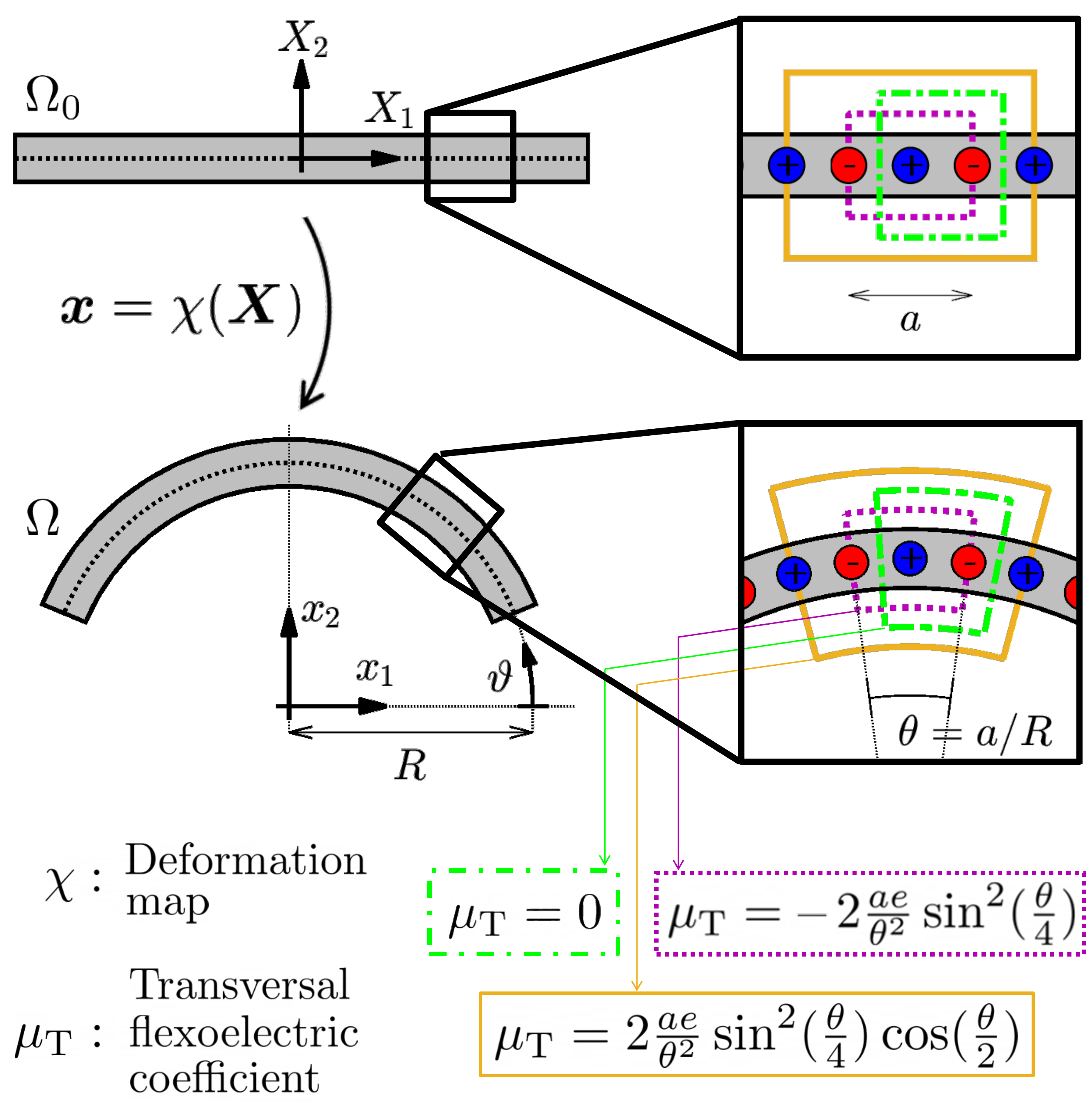}
\caption{\label{AIM} Illustration depicting the ill-defined nature of the transversal flexoelectric coefficient when the standard definition for the polarization is employed for a structure that is extended in the $X_1$-direction.}
\end{figure}

Kohn-Sham DFT calculations for $\mu_\text{T}$ at finite bending curvatures are perhaps simpler than their zero-curvature counterparts, since perturbation theory can be circumvented \cite{kalinin2008electronic, shi2018flexoelectricity, dumitricua2002curvature}. However, as illustrated in Fig.~\ref{AIM}, a fundamental issue in this context is that $\mu_\text{T}$ becomes an ill-defined quantity on employing the standard definition of polarization, i.e., dipole moment per unit volume \footnote{The Berry phase formulation is not required since the structure is finite along the direction in which the polarization is desired.}. In particular, considering a structure that is extended in the $X_1$-direction,  the value for $\mu_\text{T}$ is dependent on the choice of the unit cell in that direction. In fact, in the limiting case of the deformed unit cell encompassing the complete circle, $\mu_\text{T} = 0$  for any charge distribution,  a result that is clearly incorrect. Even for structures that are finite along the $X_1$-direction, $\mu_\text{T}$ has an artificial dependence---not attributable to edge-related effects---on the corresponding dimension of the structure, i.e., on the angle subtended by the bent structure. 

In this work, we introduce the concept of \emph{radial polarization} to overcome the ill-defined nature of transversal flexoelectric coefficient $\mu_\text{T}$. We then use this formulation to  calculate $\mu_\text{T}$ for group IV atomic monolayers along both the armchair and zigzag directions from ab initio DFT simulations.

\emph{Formulation.} Consider a deformation $\x=\boldsymbol\chi(\X)$, where the map $\boldsymbol\chi:\Omega_0\mapsto\Omega$  transforms a point with coordinates $\X=[X_1,X_2,X_3]^{\rm T}$ in the undeformed configuration $\Omega_0$ to the coordinates $\x=[x_1,x_2,x_3]^{\rm T}$ in the deformed configuration $\Omega$. The associated deformation gradient tensor $\F(\X)$ is defined as $F_{iI}:=\partial \chi_i/\partial X_I$, whose Jacobian $J:= \det(\F)$. The corresponding Green-Lagrange strain gradient tensor $\GG(\X)$ is defined as $\gg_{IJK}:= {\textstyle \frac{1}{2}} \partial \left(F_{kI}F_{kJ}\right) /\partial X_K$, where the repeated index implies summation, a notation adopted henceforth. In this finite-deformation setting, the polarization $\p(\x)$ can be expressed as \cite{Codony2020}:
\begin{gather}\label{finalpol}
p_l = \left(\epsilon_0\chi_\text{e}E_L+\mu_{LIJK}\gg_{IJK}\right) F^{-1}_{Ll} \,,
\end{gather}
where the electric field $\E(\X)$ is defined as the negative gradient of the electrostatic potential in the undeformed configuration, and $\Flexo$ the fourth-order (form II) bulk flexoelectric tensor. It can therefore be inferred that:  
\begin{gather}\label{flexofit}
\mu_{LIJK} = \left.\frac{\partial \left(p_lF_{lL}\right)}{\partial\gg_{IJK}}\right|_{\E}.
\end{gather}

On identifying $\Omega_0$ with a slab in the {$X_1$-$X_3$} plane and with some thickness in the $X_2$ direction (Fig.~\ref{AIM}), pure bending around the $X_3$ axis can be represented using the deformation: 
\begin{equation}\label{defmap}
\begin{bmatrix}
x_1 \\
x_2 \\
x_3
\end{bmatrix}
=
\boldsymbol{\chi}\left(
\begin{bmatrix}
X_1 \\
X_2 \\
X_3
\end{bmatrix}
\right)=
\begin{bmatrix}
(R+X_2)\cos\vartheta \\
(R+X_2)\sin\vartheta \\
\uplambda_3 X_3
\end{bmatrix},
\end{equation}
where $R$ is the radius of curvature, $\vartheta=\pi/2-X_1/R$, and $\uplambda_3$ is the axial stretch. The deformation gradient  and strain gradient tensors then take the form:
\begin{align} \label{Eqn:F}
\F&=\begin{bmatrix}
+(J/\uplambda_3)\sin\vartheta & \cos\vartheta & 0\\
-(J/\uplambda_3)\cos\vartheta & \sin\vartheta & 0\\
0                               & 0                      & \uplambda_3
\end{bmatrix} , \\
\GG&=\frac{J}{\uplambda_3}\begin{bmatrix}
\begin{bmatrix}
0 & 0 & 0\\
0 & 0 & 0\\
0 & 0 & 0
\end{bmatrix}
&\hspace{-.8em}
\begin{bmatrix}
1/R & 0 & 0\\
0   & 0 & 0\\
0   & 0 & 0
\end{bmatrix}
&\hspace{-.8em}
\begin{bmatrix}
0 & 0 & 0\\
0 & 0 & 0\\
0 & 0 & 0
\end{bmatrix}
\end{bmatrix} ,
\end{align}
where $J/\uplambda_3=(1+X_2/R)\approx 1$, assuming that $R$ is large  relative to the thickness of the system, which generally holds true for nanostructures.  The only component of $\GG$ that does not vanish is $\gg_{112}\approx1/R = \kappa$, where $\kappa$ is the curvature. It therefore follows from Eq.~\ref{flexofit} that the transverse flexoelectric coefficient $\flexo_\text{T} := \mu_{2112} = \partial \left(p_lF_{l2}\right)/\partial\kappa$, which can be rewritten using Eq.~\ref{Eqn:F} as:
\begin{gather}\label{flexotrans}
\flexo_\text{T} = \frac{\partial \left(\p\cdot\n\right)}{\partial \kappa} = \frac{\partial \radialpolarization}{\partial \kappa},
\end{gather}
where $ \radialpolarization:=\p\cdot\n$ is defined to be the  \emph{radial polarization}, with $\n=[\cos(\vartheta),\sin(\vartheta),0]^{\rm T}$ representing the unit vector normal to the uniformly bent structure. 

The above formulation reveals the fundamental difference between the standard and proposed definitions for the transverse flexoelectric coefficient $\flexo_\text{T}$.  Specifically, the current work suggests that $\flexo_\text{T}$ is the rate at which the radial polarization $ \radialpolarization$ changes with curvature \footnote{The definition is in agreement with reduced models for flexoelectric membranes \cite{ahmadpoor2015flexoelectricity}.}, instead of the $x_2$-component of the polarization, as assumed previously \cite{kalinin2008electronic, shi2018flexoelectricity}. In particular, the definition presented here can be viewed as a generalization of the standard one to  finite bending deformations,  agreeing in the limit $\kappa \rightarrow 0$. Indeed, the proposed formulation is applicable even to the nonlinear regime, overcoming a key limitation of the standard definition. 

In electronic structure calculations such as DFT, the radial polarization takes the form:
\begin{gather}\label{pr}
\radialpolarization=\frac{1}{\norm{\Omega}} \int_\Omega (r-R^\text{eff}) \rho (\x) \, \dd\Omega ,
\end{gather}
where $\norm{\Omega}$ denotes the volume of $\Omega$,  and the integral can be interpreted as the \emph{radial dipole moment}. Specifically, $r:=\x\cdot\n=R+X_2$ signifies the radial component of $\x$, $R^\text{eff}$ is the \emph{radial centroid} of the ions, and $\rho$ is the electron density. In obtaining the above expression, it has been assumed that the total (i.e., electrons+ions) density is charge neutral, thereby also ensuring the  invariance with respect to translations of the coordinate system. Note that $\radialpolarization$ and therefore $\flexo_\text{T}$ are independent of the choice of unit cell for structures extended in the $X_1$-direction, and do not display an artificial dependence on the corresponding width for finite structures, thereby overcoming a fundamental limitation of the standard definition.

Interestingly, the radial polarization takes the following form in  the undeformed configuration: 
\begin{gather}\label{pr2}
\radialpolarization=\frac{1}{\norm{\Omega}} \int_{\Omega_0} (X_2-X_2^\text{eff})  \rho_0 (\X) \, \dd\Omega_0 ,
\end{gather}
where $X_2^\text{eff}=R^\text{eff}-R$, and $\rho_0=J\rho$ is the nominal electron density. Therefore, the radial dipole moment in the deformed configuration $\Omega$ corresponds to the standard dipole moment along the  $X_2$-direction in the undeformed configuration $\Omega_0$.

\emph{Implementation.} The calculation of the transversal flexoelectric coefficient $\flexo_\text{T}$ requires the derivative of the radial polarization $\radialpolarization$ with respect to curvature $\kappa$, the direct evaluation of which necessitates the use of density functional perturbation theory (DFPT) \cite{gonze1997dynamical,baroni2001phonons}. Given the complexities and challenges associated with such an approach, we instead employ a numerical approximation for the derivative, which requires computing $\radialpolarization$ at multiple curvatures in the vicinity of the curvature $\kappa$ at which $\flexo_\text{T}$ is desired.

The proposed formulation for $\radialpolarization$  is not restricted by the solution scheme for the Kohn-Sham problem. It is however desirable for the chosen approach to efficiently simulate bending deformations commensurate with those found in experiments \cite{lindahl2012determination, chen2015bending, qu2019bending, han2019ultrasoft, han2019ultrasoft}.  Given the large system sizes encountered, for extended structures in particular, ab initio simulation of bending deformations is particularly challenging, even with  state-of-the-art DFT codes \cite{xu2020sparc,banerjee2018two,motamarri2020dft}. This is because DFT calculations are highly expensive, scaling cubically with system size and possessing a large prefactor, particularly when systematically improvable discretizations are used.

\begin{figure}[h!]
\includegraphics[width=.8\linewidth]{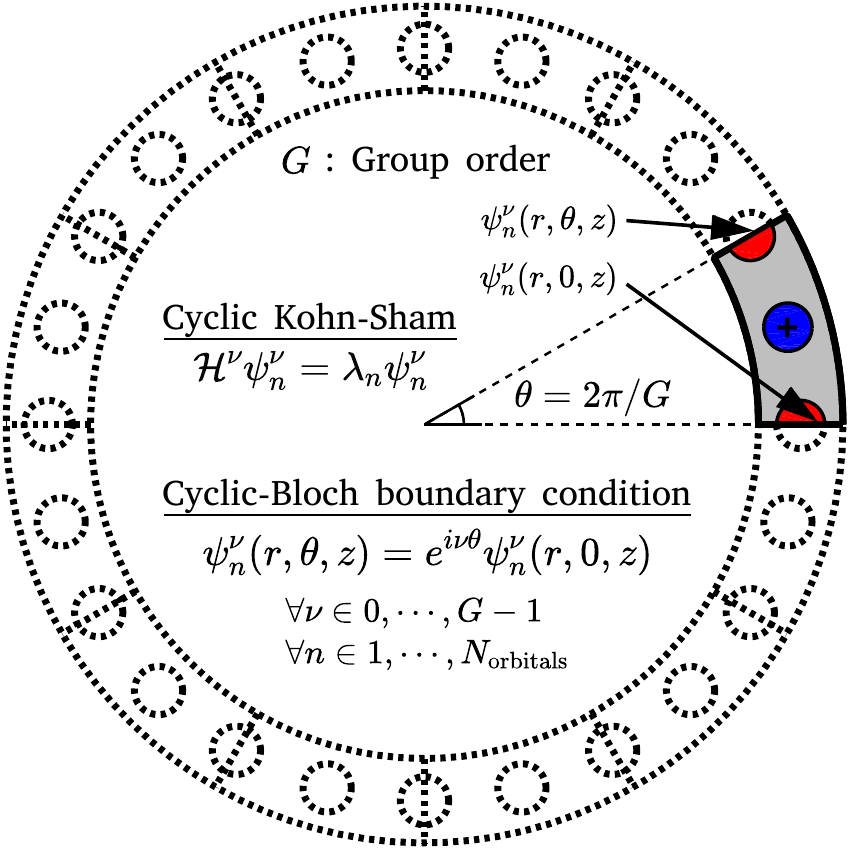}
\caption{\label{CyclicDFT} Overview of the cyclic symmetry-adapted formulation for the Kohn-Sham eigenproblem \cite{ghosh2019symmetry,Banerjee}. The Hamiltonian and  Kohn-Sham orbitals are denoted by $\mathcal{H}^{\nu}$ and $\psi_n^{\nu}$, respectively.}
\end{figure}

The calculation of  $\radialpolarization$ for structures that are extended in the $X_1$-direction requires that edge-related effects be avoided. One option is to consider a large enough structure in this direction, and use the  nearsightedness principle \cite{prodan2005nearsightedness,suryanarayana2017nearsightedness} to restrict the evaluation of $\radialpolarization$ from Eq.~\ref{pr}  to a unit cell sufficiently far from the edges. A simpler and significantly more efficient alternative, which is employed in this work, is to instead consider the complete circle for the deformed structure and exploit the cyclic symmetry present  in the system \cite{ghosh2019symmetry,Banerjee}, as illustrated in Fig.~\ref{CyclicDFT}. 

The cyclic symmetry-adapted method reduces the computations to the unit cell in the angular direction---analogous to the periodic unit cell for translational symmetry---while solving the  Kohn-Sham equations in cylindrical coordinates using the real-space finite-difference method \cite{xu2020m,xu2020sparc}.  In so doing, the computational cost scales linearly with radius of curvature, enabling tremendous savings, particularly considering the highly parallelizable nature of such calculations. This makes it the ideal tool for the study of the flexoelectric effect \cite{ghosh2019symmetry,Banerjee,Banerjee_PhD_Thesis}. Note that standard periodic boundary conditions are employed along the $x_3-$direction to account for the translational symmetry in that direction.  

\emph{Results and discussion.} We compute the transversal flexoelectric coefficient $\flexo_\text{T}$  in both armchair and zigzag directions for the group IV atomic monolayers: graphene, silicene, germanene, and stanene. This is done for two choices of exchange-correlation functional:  local density approximation (LDA) \cite{perdew1992accurate} and generalized gradient approximation (GGA) \cite{perdew1996generalized}. Optimized norm-conserving Vanderbilt (ONCV) pseudopotentials \cite{hamann2013optimized,schlipf2015optimization} are employed, whose transferability for the chosen systems has been verified by ensuring that the equilibrium monolayer structures---determined using the planewave DFT code ABINIT \cite{ABINIT}---are in good agreement with literature \cite{Novoselov10451,balendhran2015elemental}.  Curvatures of $\kappa \sim 0.19 - 0.75$ nm$^{-1}$ are considered, representative of those encountered in practice \cite{lindahl2012determination, chen2015bending, qu2019bending, han2019ultrasoft, han2019ultrasoft}. All numerical parameters are chosen so that the $\flexo_\text{T}$ are computed with an accuracy of $0.01e$. 

\begin{figure*}[htbp]
	\includegraphics[width=0.9\textwidth]{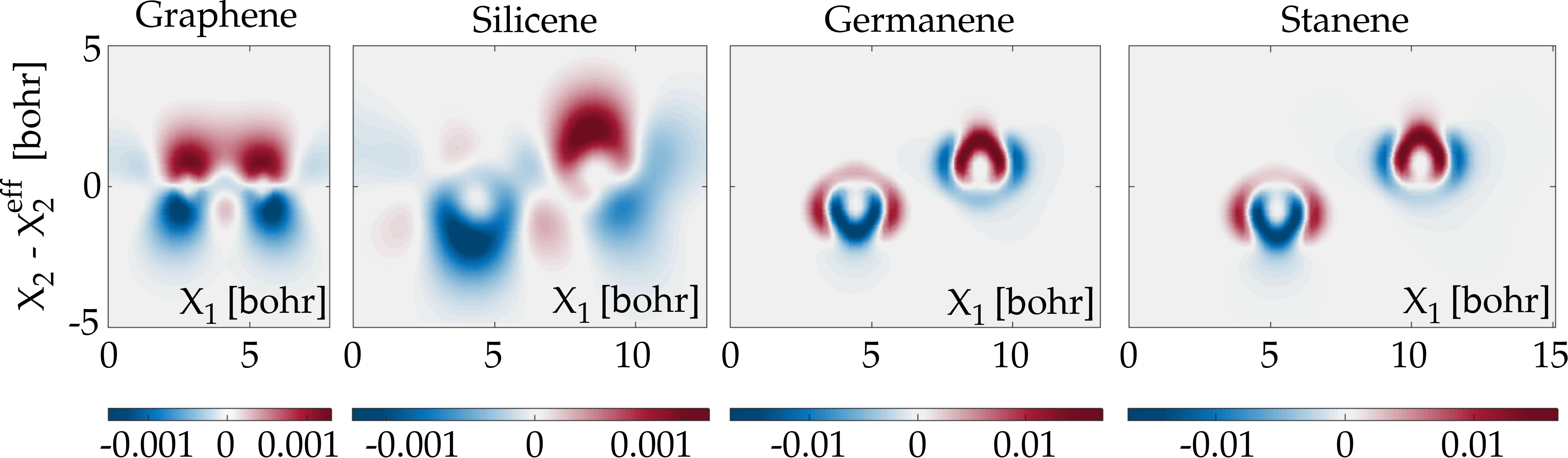}
\caption{\label{Armchair_RhoDifference} Contours of nominal electron density difference ($e$/bohr$^{3}$) between the armchair bent ($\kappa=0.19$ nm$^{-1}$) and flat atomic monolayers. The contours are in the $X_1-X_2$ plane passing through the two fundamental atoms. \vspace{-1mm} }
\end{figure*}

The values of $\flexo_\text{T}$  obtained for the group IV monolayers are presented in Table~\ref{tab:flexocoeff}. Due to the disagreement in literature over the thickness of atomic monolayers \cite{huang2006thickness}, the radial dipole moments are normalized with respect to the area instead of volume while computing the radial polarization using Eq.~\ref{pr}, i.e., the units of  $\flexo_\text{T}$ here are [$e$], rather than the conventionally used [$e$/bohr]. Note that a single curvature-independent value is listed for each entry in the table since the flexoelectric coefficients have been found to be essentially constant for the bending deformations considered here, signaling linear response for the chosen curvatures. Therefore, the values of $\flexo_\text{T}$ reported here can also be interpreted as those corresponding to the asymptotic limit of $\kappa \rightarrow 0$. 

\begin{table}[htbp]
\begin{tabular}{r|c|c|c|c|}
\cline{2-5}
\multicolumn{1}{l|}{}                    & \multicolumn{2}{c|}{\textbf{Zigzag}} & \multicolumn{2}{c|}{\textbf{Armchair}} \\ \cline{2-5} 
\multicolumn{1}{c|}{}                    & \textbf{LDA}      & \textbf{GGA}     & \textbf{LDA}       & \textbf{GGA}      \\ \hline
\multicolumn{1}{|r|}{\textbf{Graphene}}  & 0.22              & 0.22             & 0.22               & 0.22              \\ \hline
\multicolumn{1}{|r|}{\textbf{Silicene}}  & 0.19             & 0.19             & 0.19               & 0.18              \\ \hline
\multicolumn{1}{|r|}{\textbf{Germanene}} & 0.28              & 0.27             & 0.28               & 0.27              \\ \hline
\multicolumn{1}{|r|}{\textbf{Stanene}}   & 0.27              & 0.27             & 0.26               & 0.27              \\ \hline
\end{tabular}
\caption{Transversal flexoelectric coefficient $\flexo_\text{T}$ {[}$e${]} for group IV atomic monolayers. }
\label{tab:flexocoeff}
\end{table}

Notably, the results are  independent of the exchange-correlation functional, the key approximation within DFT. In addition, the nearly identical values in the zigzag and armchair directions  indicate that group IV monolayers are transersely isotropic with regards to flexoelectricity. The flexoelectric coefficients between the different materials are comparable, with germanene/stanene having the largest value ($\flexo_\text{T} \sim 0.27 e$),  silicene having the smallest ($\flexo_\text{T} \sim 0.19 e$), and graphene towards the lower end ($\flexo_\text{T}=0.22 e$). Notably, the value  for graphene is twice as large as that reported by Ref.~\cite{kalinin2008electronic}, also computed using DFT \footnote{The value is also more than two times that computed  from DFT  using the  atomic dipole model \cite{dumitricua2002curvature}, which requires an artificial partitioning of the electron density.}. The significantly smaller coefficient obtained previously can be attributed to the artificial dependence on the width, a consequence of using the standard definition of the polarization. 

To get insights into the underlying nature of the flexoelectric effect for the chosen monolayers, we plot in Fig.~\ref{Armchair_RhoDifference} the nominal electronic charge redistribution on the $X_1-X_2$ plane passing through the two fundamental atoms. For all materials, there is a net radial charge transfer that occurs from below the neutral axis to above it.  However, the plots indicate that there is a fundamental difference between graphene and the other members in its group. For graphene, bending introduces an asymmetry in the $p$-orbital overlap, leading to a rehybridization from $sp^2$ to some intermediate state between $sp^2$ and $sp^3$~\cite{nikiforov2014tight,dumitricua2002curvature,kundalwal2017strain}. However, the charge transfer in the other monolayers occurs between the two atoms and not due to the rehybridization of the orbitals within each atom. 

\emph{Concluding remarks.} In summary, we have presented a novel formulation for calculating the transversal flexoelectric coefficient of nanostructures at finite deformations from first principles. Specifically, we have introduced the concept of \emph{radial polarization} to redefine the flexoelectric coefficient, making it a  well-defined quantity for uniform bending deformations. The proposed framework has been used to calculate the coefficients for group IV atomic monolayers using DFT simulations. We have found that graphene's flexoelectric coefficient is significantly larger than that reported previously, with a charge transfer mechanism that fundamentally differs from  the other members of its group.  

The framework developed here is general and not restricted to the linear response of atomic monolayers. Therefore, it can be used  to compute the transversal flexoelectric coefficients for interesting and more complex systems, including multilayer materials such as ferroelectric perovskites, making it a worthy subject for future research. 

\emph{Acknowledgments.} 
This work was supported in part by the Generalitat de Catalunya (ICREA Academia award for excellence in research to I.A., and Grant No. 2017-SGR-1278), and the European Research Council (StG-679451 to I.A.). CIMNE is recipient of a Severo Ochoa Award of Excellence from the MINECO. P.S. gratefully   acknowledges   the   support   of the U.S. National Science Foundation (CAREER-1553212).  The authors acknowledge discussions with Shashikant Kumar and his help with some simulations. P.S. acknowledges discussions with Amartya Banerjee prior to starting this work. \vspace{-1mm}

\bibliographystyle{apsrev4-1}

\begin{thebibliography}{46}%
\makeatletter
\providecommand \@ifxundefined [1]{%
 \@ifx{#1\undefined}
}%
\providecommand \@ifnum [1]{%
 \ifnum #1\expandafter \@firstoftwo
 \else \expandafter \@secondoftwo
 \fi
}%
\providecommand \@ifx [1]{%
 \ifx #1\expandafter \@firstoftwo
 \else \expandafter \@secondoftwo
 \fi
}%
\providecommand \natexlab [1]{#1}%
\providecommand \enquote  [1]{``#1''}%
\providecommand \bibnamefont  [1]{#1}%
\providecommand \bibfnamefont [1]{#1}%
\providecommand \citenamefont [1]{#1}%
\providecommand \href@noop [0]{\@secondoftwo}%
\providecommand \href [0]{\begingroup \@sanitize@url \@href}%
\providecommand \@href[1]{\@@startlink{#1}\@@href}%
\providecommand \@@href[1]{\endgroup#1\@@endlink}%
\providecommand \@sanitize@url [0]{\catcode `\\12\catcode `\$12\catcode
  `\&12\catcode `\#12\catcode `\^12\catcode `\_12\catcode `\%12\relax}%
\providecommand \@@startlink[1]{}%
\providecommand \@@endlink[0]{}%
\providecommand \url  [0]{\begingroup\@sanitize@url \@url }%
\providecommand \@url [1]{\endgroup\@href {#1}{\urlprefix }}%
\providecommand \urlprefix  [0]{URL }%
\providecommand \Eprint [0]{\href }%
\providecommand \doibase [0]{http://dx.doi.org/}%
\providecommand \selectlanguage [0]{\@gobble}%
\providecommand \bibinfo  [0]{\@secondoftwo}%
\providecommand \bibfield  [0]{\@secondoftwo}%
\providecommand \translation [1]{[#1]}%
\providecommand \BibitemOpen [0]{}%
\providecommand \bibitemStop [0]{}%
\providecommand \bibitemNoStop [0]{.\EOS\space}%
\providecommand \EOS [0]{\spacefactor3000\relax}%
\providecommand \BibitemShut  [1]{\csname bibitem#1\endcsname}%
\let\auto@bib@innerbib\@empty
\bibitem [{\citenamefont {Tagantsev}(1991)}]{tagantsev1991electric}%
  \BibitemOpen
  \bibfield  {author} {\bibinfo {author} {\bibfnamefont {A.~K.}\ \bibnamefont
  {Tagantsev}},\ }\href {https://doi.org/10.1080/01411599108213201} {\bibfield
  {journal} {\bibinfo  {journal} {Phase Transit.}\ }\textbf {\bibinfo {volume}
  {35}},\ \bibinfo {pages} {119} (\bibinfo {year} {1991})}\BibitemShut
  {NoStop}%
\bibitem [{\citenamefont {Yudin}\ and\ \citenamefont
  {Tagantsev}(2013)}]{yudin2013fundamentals}%
  \BibitemOpen
  \bibfield  {author} {\bibinfo {author} {\bibfnamefont {P.}~\bibnamefont
  {Yudin}}\ and\ \bibinfo {author} {\bibfnamefont {A.}~\bibnamefont
  {Tagantsev}},\ }\href
  {https://doi.org/10.1088/0957-4484\%2F24\%2F43\%2F432001} {\bibfield
  {journal} {\bibinfo  {journal} {Nanotechnology}\ }\textbf {\bibinfo {volume}
  {24}},\ \bibinfo {pages} {432001} (\bibinfo {year} {2013})}\BibitemShut
  {NoStop}%
\bibitem [{\citenamefont {Zubko}\ \emph {et~al.}(2013)\citenamefont {Zubko},
  \citenamefont {Catalan},\ and\ \citenamefont
  {Tagantsev}}]{zubko2013flexoelectric}%
  \BibitemOpen
  \bibfield  {author} {\bibinfo {author} {\bibfnamefont {P.}~\bibnamefont
  {Zubko}}, \bibinfo {author} {\bibfnamefont {G.}~\bibnamefont {Catalan}}, \
  and\ \bibinfo {author} {\bibfnamefont {A.~K.}\ \bibnamefont {Tagantsev}},\
  }\href {https://doi.org/10.1146/annurev-matsci-071312-121634} {\bibfield
  {journal} {\bibinfo  {journal} {Annu. Rev. Mater. Sci.}\ }\textbf {\bibinfo
  {volume} {43}} (\bibinfo {year} {2013})}\BibitemShut {NoStop}%
\bibitem [{\citenamefont {Nguyen}\ \emph {et~al.}(2013)\citenamefont {Nguyen},
  \citenamefont {Mao}, \citenamefont {Yeh}, \citenamefont {Purohit},\ and\
  \citenamefont {McAlpine}}]{nguyen2013nanoscale}%
  \BibitemOpen
  \bibfield  {author} {\bibinfo {author} {\bibfnamefont {T.~D.}\ \bibnamefont
  {Nguyen}}, \bibinfo {author} {\bibfnamefont {S.}~\bibnamefont {Mao}},
  \bibinfo {author} {\bibfnamefont {Y.-W.}\ \bibnamefont {Yeh}}, \bibinfo
  {author} {\bibfnamefont {P.~K.}\ \bibnamefont {Purohit}}, \ and\ \bibinfo
  {author} {\bibfnamefont {M.~C.}\ \bibnamefont {McAlpine}},\ }\href
  {https://doi.org/10.1002/adma.201203852} {\bibfield  {journal} {\bibinfo
  {journal} {Adv. Mat.}\ }\textbf {\bibinfo {volume} {25}},\ \bibinfo {pages}
  {946} (\bibinfo {year} {2013})}\BibitemShut {NoStop}%
\bibitem [{\citenamefont {Ahmadpoor}\ and\ \citenamefont
  {Sharma}(2015)}]{ahmadpoor2015flexoelectricity}%
  \BibitemOpen
  \bibfield  {author} {\bibinfo {author} {\bibfnamefont {F.}~\bibnamefont
  {Ahmadpoor}}\ and\ \bibinfo {author} {\bibfnamefont {P.}~\bibnamefont
  {Sharma}},\ }\href {https://doi.org/10.1039/C5NR04722F} {\bibfield  {journal}
  {\bibinfo  {journal} {Nanoscale}\ }\textbf {\bibinfo {volume} {7}},\ \bibinfo
  {pages} {16555} (\bibinfo {year} {2015})}\BibitemShut {NoStop}%
\bibitem [{\citenamefont {Krichen}\ and\ \citenamefont
  {Sharma}(2016)}]{krichen2016flexoelectricity}%
  \BibitemOpen
  \bibfield  {author} {\bibinfo {author} {\bibfnamefont {S.}~\bibnamefont
  {Krichen}}\ and\ \bibinfo {author} {\bibfnamefont {P.}~\bibnamefont
  {Sharma}},\ }\href {https://doi.org/10.1115/1.4032378} {\bibfield  {journal}
  {\bibinfo  {journal} {J. Appl. Mech.}\ }\textbf {\bibinfo {volume} {83}}
  (\bibinfo {year} {2016})}\BibitemShut {NoStop}%
\bibitem [{\citenamefont {Wang}\ \emph {et~al.}(2019)\citenamefont {Wang},
  \citenamefont {Gu}, \citenamefont {Zhang},\ and\ \citenamefont
  {Chen}}]{wang2019flexoelectricity}%
  \BibitemOpen
  \bibfield  {author} {\bibinfo {author} {\bibfnamefont {B.}~\bibnamefont
  {Wang}}, \bibinfo {author} {\bibfnamefont {Y.}~\bibnamefont {Gu}}, \bibinfo
  {author} {\bibfnamefont {S.}~\bibnamefont {Zhang}}, \ and\ \bibinfo {author}
  {\bibfnamefont {L.-Q.}\ \bibnamefont {Chen}},\ }\href
  {https://doi.org/10.1016/j.pmatsci.2019.05.003} {\bibfield  {journal}
  {\bibinfo  {journal} {Prog. Mater. Sci.}\ }\textbf {\bibinfo {volume}
  {106}},\ \bibinfo {pages} {100570} (\bibinfo {year} {2019})}\BibitemShut
  {NoStop}%
\bibitem [{\citenamefont {Hong}\ \emph {et~al.}(2010)\citenamefont {Hong},
  \citenamefont {Catalan}, \citenamefont {Scott},\ and\ \citenamefont
  {Artacho}}]{hong2010flexoelectricity}%
  \BibitemOpen
  \bibfield  {author} {\bibinfo {author} {\bibfnamefont {J.}~\bibnamefont
  {Hong}}, \bibinfo {author} {\bibfnamefont {G.}~\bibnamefont {Catalan}},
  \bibinfo {author} {\bibfnamefont {J.}~\bibnamefont {Scott}}, \ and\ \bibinfo
  {author} {\bibfnamefont {E.}~\bibnamefont {Artacho}},\ }\href
  {https://doi.org/10.1088/0953-8984/22/11/112201} {\bibfield  {journal}
  {\bibinfo  {journal} {J. Phys. Condens. Matter.}\ }\textbf {\bibinfo {volume}
  {22}},\ \bibinfo {pages} {112201} (\bibinfo {year} {2010})}\BibitemShut
  {NoStop}%
\bibitem [{\citenamefont {Kohn}\ and\ \citenamefont
  {Sham}(1965)}]{KohnSham_DFT}%
  \BibitemOpen
  \bibfield  {author} {\bibinfo {author} {\bibfnamefont {W.}~\bibnamefont
  {Kohn}}\ and\ \bibinfo {author} {\bibfnamefont {L.~J.}\ \bibnamefont
  {Sham}},\ }\href {https://doi.org/10.1103/PhysRev.140.A1133} {\bibfield
  {journal} {\bibinfo  {journal} {Phys. Rev.}\ }\textbf {\bibinfo {volume}
  {140}},\ \bibinfo {pages} {A1133} (\bibinfo {year} {1965})}\BibitemShut
  {NoStop}%
\bibitem [{\citenamefont {Hong}\ and\ \citenamefont
  {Vanderbilt}(2011)}]{hong2011first}%
  \BibitemOpen
  \bibfield  {author} {\bibinfo {author} {\bibfnamefont {J.}~\bibnamefont
  {Hong}}\ and\ \bibinfo {author} {\bibfnamefont {D.}~\bibnamefont
  {Vanderbilt}},\ }\href {https://doi.org/10.1103/PhysRevB.88.174107}
  {\bibfield  {journal} {\bibinfo  {journal} {Phys. Rev. B}\ }\textbf {\bibinfo
  {volume} {84}},\ \bibinfo {pages} {180101(R)} (\bibinfo {year}
  {2011})}\BibitemShut {NoStop}%
\bibitem [{\citenamefont {Hong}\ and\ \citenamefont
  {Vanderbilt}(2013)}]{hong2013first}%
  \BibitemOpen
  \bibfield  {author} {\bibinfo {author} {\bibfnamefont {J.}~\bibnamefont
  {Hong}}\ and\ \bibinfo {author} {\bibfnamefont {D.}~\bibnamefont
  {Vanderbilt}},\ }\href {https://doi.org/10.1103/PhysRevB.88.174107}
  {\bibfield  {journal} {\bibinfo  {journal} {Phys. Rev. B}\ }\textbf {\bibinfo
  {volume} {88}},\ \bibinfo {pages} {174107} (\bibinfo {year}
  {2013})}\BibitemShut {NoStop}%
\bibitem [{\citenamefont {Stengel}(2013)}]{stengel2013flexoelectricity}%
  \BibitemOpen
  \bibfield  {author} {\bibinfo {author} {\bibfnamefont {M.}~\bibnamefont
  {Stengel}},\ }\href {https://doi.org/10.1103/PhysRevB.88.174106} {\bibfield
  {journal} {\bibinfo  {journal} {Phys. Rev. B}\ }\textbf {\bibinfo {volume}
  {88}},\ \bibinfo {pages} {174106} (\bibinfo {year} {2013})}\BibitemShut
  {NoStop}%
\bibitem [{\citenamefont {Stengel}(2014)}]{stengel2014surface}%
  \BibitemOpen
  \bibfield  {author} {\bibinfo {author} {\bibfnamefont {M.}~\bibnamefont
  {Stengel}},\ }\href {https://doi.org/10.1103/PhysRevB.90.201112} {\bibfield
  {journal} {\bibinfo  {journal} {Phys. Rev. B}\ }\textbf {\bibinfo {volume}
  {90}},\ \bibinfo {pages} {201112(R)} (\bibinfo {year} {2014})}\BibitemShut
  {NoStop}%
\bibitem [{\citenamefont {Dreyer}\ \emph {et~al.}(2018)\citenamefont {Dreyer},
  \citenamefont {Stengel},\ and\ \citenamefont
  {Vanderbilt}}]{dreyer2018current}%
  \BibitemOpen
  \bibfield  {author} {\bibinfo {author} {\bibfnamefont {C.~E.}\ \bibnamefont
  {Dreyer}}, \bibinfo {author} {\bibfnamefont {M.}~\bibnamefont {Stengel}}, \
  and\ \bibinfo {author} {\bibfnamefont {D.}~\bibnamefont {Vanderbilt}},\
  }\href {https://doi.org/10.1103/PhysRevB.98.075153} {\bibfield  {journal}
  {\bibinfo  {journal} {Phys. Rev. B}\ }\textbf {\bibinfo {volume} {98}},\
  \bibinfo {pages} {075153} (\bibinfo {year} {2018})}\BibitemShut {NoStop}%
\bibitem [{\citenamefont {Lindahl}\ \emph {et~al.}(2012)\citenamefont
  {Lindahl}, \citenamefont {Midtvedt}, \citenamefont {Svensson}, \citenamefont
  {Nerushev}, \citenamefont {Lindvall}, \citenamefont {Isacsson},\ and\
  \citenamefont {Campbell}}]{lindahl2012determination}%
  \BibitemOpen
  \bibfield  {author} {\bibinfo {author} {\bibfnamefont {N.}~\bibnamefont
  {Lindahl}}, \bibinfo {author} {\bibfnamefont {D.}~\bibnamefont {Midtvedt}},
  \bibinfo {author} {\bibfnamefont {J.}~\bibnamefont {Svensson}}, \bibinfo
  {author} {\bibfnamefont {O.~A.}\ \bibnamefont {Nerushev}}, \bibinfo {author}
  {\bibfnamefont {N.}~\bibnamefont {Lindvall}}, \bibinfo {author}
  {\bibfnamefont {A.}~\bibnamefont {Isacsson}}, \ and\ \bibinfo {author}
  {\bibfnamefont {E.~E.}\ \bibnamefont {Campbell}},\ }\href
  {https://doi.org/10.1021/nl301080v} {\bibfield  {journal} {\bibinfo
  {journal} {Nano Lett.}\ }\textbf {\bibinfo {volume} {12}},\ \bibinfo {pages}
  {3526} (\bibinfo {year} {2012})}\BibitemShut {NoStop}%
\bibitem [{\citenamefont {Chen}\ \emph {et~al.}(2015)\citenamefont {Chen},
  \citenamefont {Yi},\ and\ \citenamefont {Ke}}]{chen2015bending}%
  \BibitemOpen
  \bibfield  {author} {\bibinfo {author} {\bibfnamefont {X.}~\bibnamefont
  {Chen}}, \bibinfo {author} {\bibfnamefont {C.}~\bibnamefont {Yi}}, \ and\
  \bibinfo {author} {\bibfnamefont {C.}~\bibnamefont {Ke}},\ }\href
  {https://doi.org/10.1063/1.4915075} {\bibfield  {journal} {\bibinfo
  {journal} {Appl. Phys. Lett.}\ }\textbf {\bibinfo {volume} {106}},\ \bibinfo
  {pages} {101907} (\bibinfo {year} {2015})}\BibitemShut {NoStop}%
\bibitem [{\citenamefont {Qu}\ \emph {et~al.}(2019)\citenamefont {Qu},
  \citenamefont {Bagchi}, \citenamefont {Chen}, \citenamefont {Chew},\ and\
  \citenamefont {Ke}}]{qu2019bending}%
  \BibitemOpen
  \bibfield  {author} {\bibinfo {author} {\bibfnamefont {W.}~\bibnamefont
  {Qu}}, \bibinfo {author} {\bibfnamefont {S.}~\bibnamefont {Bagchi}}, \bibinfo
  {author} {\bibfnamefont {X.}~\bibnamefont {Chen}}, \bibinfo {author}
  {\bibfnamefont {H.~B.}\ \bibnamefont {Chew}}, \ and\ \bibinfo {author}
  {\bibfnamefont {C.}~\bibnamefont {Ke}},\ }\href
  {https://doi.org/10.1038/s41563-019-0529-7} {\bibfield  {journal} {\bibinfo
  {journal} {J. Phys. D}\ }\textbf {\bibinfo {volume} {52}},\ \bibinfo {pages}
  {465301} (\bibinfo {year} {2019})}\BibitemShut {NoStop}%
\bibitem [{\citenamefont {Han}\ \emph {et~al.}(2019)\citenamefont {Han},
  \citenamefont {Yu}, \citenamefont {Annevelink}, \citenamefont {Son},
  \citenamefont {Kang}, \citenamefont {Watanabe}, \citenamefont {Taniguchi},
  \citenamefont {Ertekin}, \citenamefont {Huang},\ and\ \citenamefont {van~der
  Zande}}]{han2019ultrasoft}%
  \BibitemOpen
  \bibfield  {author} {\bibinfo {author} {\bibfnamefont {E.}~\bibnamefont
  {Han}}, \bibinfo {author} {\bibfnamefont {J.}~\bibnamefont {Yu}}, \bibinfo
  {author} {\bibfnamefont {E.}~\bibnamefont {Annevelink}}, \bibinfo {author}
  {\bibfnamefont {J.}~\bibnamefont {Son}}, \bibinfo {author} {\bibfnamefont
  {D.~A.}\ \bibnamefont {Kang}}, \bibinfo {author} {\bibfnamefont
  {K.}~\bibnamefont {Watanabe}}, \bibinfo {author} {\bibfnamefont
  {T.}~\bibnamefont {Taniguchi}}, \bibinfo {author} {\bibfnamefont
  {E.}~\bibnamefont {Ertekin}}, \bibinfo {author} {\bibfnamefont {P.~Y.}\
  \bibnamefont {Huang}}, \ and\ \bibinfo {author} {\bibfnamefont {A.~M.}\
  \bibnamefont {van~der Zande}},\ }\href
  {https://doi.org/10.1038/s41563-019-0529-7} {\bibfield  {journal} {\bibinfo
  {journal} {Nat. Mater.}\ ,\ \bibinfo {pages} {1}} (\bibinfo {year}
  {2019})}\BibitemShut {NoStop}%
\bibitem [{\citenamefont {Kalinin}\ and\ \citenamefont
  {Meunier}(2008)}]{kalinin2008electronic}%
  \BibitemOpen
  \bibfield  {author} {\bibinfo {author} {\bibfnamefont {S.~V.}\ \bibnamefont
  {Kalinin}}\ and\ \bibinfo {author} {\bibfnamefont {V.}~\bibnamefont
  {Meunier}},\ }\href {https://doi.org/10.1103/PhysRevB.77.033403} {\bibfield
  {journal} {\bibinfo  {journal} {Phys. Rev. B}\ }\textbf {\bibinfo {volume}
  {77}},\ \bibinfo {pages} {033403} (\bibinfo {year} {2008})}\BibitemShut
  {NoStop}%
\bibitem [{\citenamefont {Shi}\ \emph {et~al.}(2018)\citenamefont {Shi},
  \citenamefont {Guo}, \citenamefont {Zhang},\ and\ \citenamefont
  {Guo}}]{shi2018flexoelectricity}%
  \BibitemOpen
  \bibfield  {author} {\bibinfo {author} {\bibfnamefont {W.}~\bibnamefont
  {Shi}}, \bibinfo {author} {\bibfnamefont {Y.}~\bibnamefont {Guo}}, \bibinfo
  {author} {\bibfnamefont {Z.}~\bibnamefont {Zhang}}, \ and\ \bibinfo {author}
  {\bibfnamefont {W.}~\bibnamefont {Guo}},\ }\href
  {https://doi.org/10.1021/acs.jpclett.8b03325} {\bibfield  {journal} {\bibinfo
   {journal} {J. Phys. Chem. Lett.}\ }\textbf {\bibinfo {volume} {9}},\
  \bibinfo {pages} {6841} (\bibinfo {year} {2018})}\BibitemShut {NoStop}%
\bibitem [{\citenamefont {Dumitric{\u{a}}}\ \emph {et~al.}(2002)\citenamefont
  {Dumitric{\u{a}}}, \citenamefont {Landis},\ and\ \citenamefont
  {Yakobson}}]{dumitricua2002curvature}%
  \BibitemOpen
  \bibfield  {author} {\bibinfo {author} {\bibfnamefont {T.}~\bibnamefont
  {Dumitric{\u{a}}}}, \bibinfo {author} {\bibfnamefont {C.~M.}\ \bibnamefont
  {Landis}}, \ and\ \bibinfo {author} {\bibfnamefont {B.~I.}\ \bibnamefont
  {Yakobson}},\ }\href {https://doi.org/10.1016/S0009-2614(02)00820-5}
  {\bibfield  {journal} {\bibinfo  {journal} {Chem. Phys. Lett.}\ }\textbf
  {\bibinfo {volume} {360}},\ \bibinfo {pages} {182} (\bibinfo {year}
  {2002})}\BibitemShut {NoStop}%
\bibitem [{Note1()}]{Note1}%
  \BibitemOpen
  \bibinfo {note} {The Berry phase formulation is not required since the
  structure is finite along the direction in which the polarization is
  desired.}\BibitemShut {Stop}%
\bibitem [{\citenamefont {Codony}\ \emph {et~al.}(2020)\citenamefont {Codony},
  \citenamefont {Gupta}, \citenamefont {Marco},\ and\ \citenamefont
  {Arias}}]{Codony2020}%
  \BibitemOpen
  \bibfield  {author} {\bibinfo {author} {\bibfnamefont {D.}~\bibnamefont
  {Codony}}, \bibinfo {author} {\bibfnamefont {P.}~\bibnamefont {Gupta}},
  \bibinfo {author} {\bibfnamefont {O.}~\bibnamefont {Marco}}, \ and\ \bibinfo
  {author} {\bibfnamefont {I.}~\bibnamefont {Arias}},\ }\href
  {https://arxiv.org/abs/2008.09045} {\bibfield  {journal} {\bibinfo  {journal}
  {arXiv preprint arXiv:2008.09045}\ } (\bibinfo {year} {2020})}\BibitemShut
  {NoStop}%
\bibitem [{Note2()}]{Note2}%
  \BibitemOpen
  \bibinfo {note} {The definition is in agreement with reduced models for
  flexoelectric membranes \cite {ahmadpoor2015flexoelectricity}.}\BibitemShut
  {Stop}%
\bibitem [{\citenamefont {Gonze}\ and\ \citenamefont
  {Lee}(1997)}]{gonze1997dynamical}%
  \BibitemOpen
  \bibfield  {author} {\bibinfo {author} {\bibfnamefont {X.}~\bibnamefont
  {Gonze}}\ and\ \bibinfo {author} {\bibfnamefont {C.}~\bibnamefont {Lee}},\
  }\href {https://doi.org/10.1103/PhysRevB.55.10355} {\bibfield  {journal}
  {\bibinfo  {journal} {Phys. Rev. B}\ }\textbf {\bibinfo {volume} {55}},\
  \bibinfo {pages} {10355} (\bibinfo {year} {1997})}\BibitemShut {NoStop}%
\bibitem [{\citenamefont {Baroni}\ \emph {et~al.}(2001)\citenamefont {Baroni},
  \citenamefont {De~Gironcoli}, \citenamefont {Dal~Corso},\ and\ \citenamefont
  {Giannozzi}}]{baroni2001phonons}%
  \BibitemOpen
  \bibfield  {author} {\bibinfo {author} {\bibfnamefont {S.}~\bibnamefont
  {Baroni}}, \bibinfo {author} {\bibfnamefont {S.}~\bibnamefont
  {De~Gironcoli}}, \bibinfo {author} {\bibfnamefont {A.}~\bibnamefont
  {Dal~Corso}}, \ and\ \bibinfo {author} {\bibfnamefont {P.}~\bibnamefont
  {Giannozzi}},\ }\href {https://doi.org/10.1103/RevModPhys.73.515} {\bibfield
  {journal} {\bibinfo  {journal} {Rev. Mod. Phys.}\ }\textbf {\bibinfo {volume}
  {73}},\ \bibinfo {pages} {515} (\bibinfo {year} {2001})}\BibitemShut
  {NoStop}%
\bibitem [{\citenamefont {Xu}\ \emph {et~al.}(2020{\natexlab{a}})\citenamefont
  {Xu}, \citenamefont {Sharma}, \citenamefont {Comer}, \citenamefont {Huang},
  \citenamefont {Chow}, \citenamefont {Medford}, \citenamefont {Pask},\ and\
  \citenamefont {Suryanarayana}}]{xu2020sparc}%
  \BibitemOpen
  \bibfield  {author} {\bibinfo {author} {\bibfnamefont {Q.}~\bibnamefont
  {Xu}}, \bibinfo {author} {\bibfnamefont {A.}~\bibnamefont {Sharma}}, \bibinfo
  {author} {\bibfnamefont {B.}~\bibnamefont {Comer}}, \bibinfo {author}
  {\bibfnamefont {H.}~\bibnamefont {Huang}}, \bibinfo {author} {\bibfnamefont
  {E.}~\bibnamefont {Chow}}, \bibinfo {author} {\bibfnamefont {A.~J.}\
  \bibnamefont {Medford}}, \bibinfo {author} {\bibfnamefont {J.~E.}\
  \bibnamefont {Pask}}, \ and\ \bibinfo {author} {\bibfnamefont
  {P.}~\bibnamefont {Suryanarayana}},\ }\href
  {https://arxiv.org/abs/2005.10431} {\bibfield  {journal} {\bibinfo  {journal}
  {arXiv preprint arXiv:2005.10431}\ } (\bibinfo {year}
  {2020}{\natexlab{a}})}\BibitemShut {NoStop}%
\bibitem [{\citenamefont {Banerjee}\ \emph {et~al.}(2018)\citenamefont
  {Banerjee}, \citenamefont {Lin}, \citenamefont {Suryanarayana}, \citenamefont
  {Yang},\ and\ \citenamefont {Pask}}]{banerjee2018two}%
  \BibitemOpen
  \bibfield  {author} {\bibinfo {author} {\bibfnamefont {A.~S.}\ \bibnamefont
  {Banerjee}}, \bibinfo {author} {\bibfnamefont {L.}~\bibnamefont {Lin}},
  \bibinfo {author} {\bibfnamefont {P.}~\bibnamefont {Suryanarayana}}, \bibinfo
  {author} {\bibfnamefont {C.}~\bibnamefont {Yang}}, \ and\ \bibinfo {author}
  {\bibfnamefont {J.~E.}\ \bibnamefont {Pask}},\ }\href
  {https://doi.org/10.1021/acs.jctc.7b01243} {\bibfield  {journal} {\bibinfo
  {journal} {J. Chem. Theory Comput.}\ }\textbf {\bibinfo {volume} {14}},\
  \bibinfo {pages} {2930} (\bibinfo {year} {2018})}\BibitemShut {NoStop}%
\bibitem [{\citenamefont {Motamarri}\ \emph {et~al.}(2020)\citenamefont
  {Motamarri}, \citenamefont {Das}, \citenamefont {Rudraraju}, \citenamefont
  {Ghosh}, \citenamefont {Davydov},\ and\ \citenamefont
  {Gavini}}]{motamarri2020dft}%
  \BibitemOpen
  \bibfield  {author} {\bibinfo {author} {\bibfnamefont {P.}~\bibnamefont
  {Motamarri}}, \bibinfo {author} {\bibfnamefont {S.}~\bibnamefont {Das}},
  \bibinfo {author} {\bibfnamefont {S.}~\bibnamefont {Rudraraju}}, \bibinfo
  {author} {\bibfnamefont {K.}~\bibnamefont {Ghosh}}, \bibinfo {author}
  {\bibfnamefont {D.}~\bibnamefont {Davydov}}, \ and\ \bibinfo {author}
  {\bibfnamefont {V.}~\bibnamefont {Gavini}},\ }\href
  {https://doi.org/10.1016/j.cpc.2019.07.016} {\bibfield  {journal} {\bibinfo
  {journal} {Comput. Phys. Commun.}\ }\textbf {\bibinfo {volume} {246}},\
  \bibinfo {pages} {106853} (\bibinfo {year} {2020})}\BibitemShut {NoStop}%
\bibitem [{\citenamefont {Prodan}\ and\ \citenamefont
  {Kohn}(2005)}]{prodan2005nearsightedness}%
  \BibitemOpen
  \bibfield  {author} {\bibinfo {author} {\bibfnamefont {E.}~\bibnamefont
  {Prodan}}\ and\ \bibinfo {author} {\bibfnamefont {W.}~\bibnamefont {Kohn}},\
  }\href {\doibase 10.1073/pnas.0505436102} {\bibfield  {journal} {\bibinfo
  {journal} {Proc. Natl. Acad. Sci. U. S. A.}\ }\textbf {\bibinfo {volume}
  {102}},\ \bibinfo {pages} {11635} (\bibinfo {year} {2005})}\BibitemShut
  {NoStop}%
\bibitem [{\citenamefont
  {Suryanarayana}(2017)}]{suryanarayana2017nearsightedness}%
  \BibitemOpen
  \bibfield  {author} {\bibinfo {author} {\bibfnamefont {P.}~\bibnamefont
  {Suryanarayana}},\ }\href {\doibase 10.1016/j.cplett.2017.04.095} {\bibfield
  {journal} {\bibinfo  {journal} {Chem. Phys. Lett.}\ }\textbf {\bibinfo
  {volume} {679}},\ \bibinfo {pages} {146} (\bibinfo {year}
  {2017})}\BibitemShut {NoStop}%
\bibitem [{\citenamefont {Ghosh}\ \emph {et~al.}(2019)\citenamefont {Ghosh},
  \citenamefont {Banerjee},\ and\ \citenamefont
  {Suryanarayana}}]{ghosh2019symmetry}%
  \BibitemOpen
  \bibfield  {author} {\bibinfo {author} {\bibfnamefont {S.}~\bibnamefont
  {Ghosh}}, \bibinfo {author} {\bibfnamefont {A.~S.}\ \bibnamefont {Banerjee}},
  \ and\ \bibinfo {author} {\bibfnamefont {P.}~\bibnamefont {Suryanarayana}},\
  }\href {https://doi.org/10.1103/PhysRevB.100.125143} {\bibfield  {journal}
  {\bibinfo  {journal} {Phys. Rev. B}\ }\textbf {\bibinfo {volume} {100}},\
  \bibinfo {pages} {125143} (\bibinfo {year} {2019})}\BibitemShut {NoStop}%
\bibitem [{\citenamefont {Banerjee}\ and\ \citenamefont
  {Suryanarayana}(2016)}]{Banerjee}%
  \BibitemOpen
  \bibfield  {author} {\bibinfo {author} {\bibfnamefont {A.~S.}\ \bibnamefont
  {Banerjee}}\ and\ \bibinfo {author} {\bibfnamefont {P.}~\bibnamefont
  {Suryanarayana}},\ }\href {https://doi.org/10.1016/j.jmps.2016.08.007}
  {\bibfield  {journal} {\bibinfo  {journal} {J. Mech. Phys. Solids}\ }\textbf
  {\bibinfo {volume} {96}},\ \bibinfo {pages} {605} (\bibinfo {year}
  {2016})}\BibitemShut {NoStop}%
\bibitem [{\citenamefont {Xu}\ \emph {et~al.}(2020{\natexlab{b}})\citenamefont
  {Xu}, \citenamefont {Sharma},\ and\ \citenamefont {Suryanarayana}}]{xu2020m}%
  \BibitemOpen
  \bibfield  {author} {\bibinfo {author} {\bibfnamefont {Q.}~\bibnamefont
  {Xu}}, \bibinfo {author} {\bibfnamefont {A.}~\bibnamefont {Sharma}}, \ and\
  \bibinfo {author} {\bibfnamefont {P.}~\bibnamefont {Suryanarayana}},\ }\href
  {https://doi.org/10.1016/j.softx.2020.100423} {\bibfield  {journal} {\bibinfo
   {journal} {SoftwareX}\ }\textbf {\bibinfo {volume} {11}},\ \bibinfo {pages}
  {100423} (\bibinfo {year} {2020}{\natexlab{b}})}\BibitemShut {NoStop}%
\bibitem [{\citenamefont {Banerjee}(2013)}]{Banerjee_PhD_Thesis}%
  \BibitemOpen
  \bibfield  {author} {\bibinfo {author} {\bibfnamefont {A.~S.}\ \bibnamefont
  {Banerjee}},\ }\emph {\bibinfo {title} {Density Functional Methods for
  {O}bjective {S}tructures: Theory and Simulation Schemes}},\ \href@noop {}
  {Ph.D. thesis},\ \bibinfo  {school} {University of Minnesota, Minneapolis},
  \bibinfo {address} {Minneapolis, MN} (\bibinfo {year} {2013})\BibitemShut
  {NoStop}%
\bibitem [{\citenamefont {Perdew}\ and\ \citenamefont
  {Wang}(1992)}]{perdew1992accurate}%
  \BibitemOpen
  \bibfield  {author} {\bibinfo {author} {\bibfnamefont {J.~P.}\ \bibnamefont
  {Perdew}}\ and\ \bibinfo {author} {\bibfnamefont {Y.}~\bibnamefont {Wang}},\
  }\href {https://doi.org/10.1103/PhysRevB.45.13244} {\bibfield  {journal}
  {\bibinfo  {journal} {Phys. Rev. B}\ }\textbf {\bibinfo {volume} {45}},\
  \bibinfo {pages} {13244} (\bibinfo {year} {1992})}\BibitemShut {NoStop}%
\bibitem [{\citenamefont {Perdew}\ \emph {et~al.}(1996)\citenamefont {Perdew},
  \citenamefont {Burke},\ and\ \citenamefont
  {Ernzerhof}}]{perdew1996generalized}%
  \BibitemOpen
  \bibfield  {author} {\bibinfo {author} {\bibfnamefont {J.~P.}\ \bibnamefont
  {Perdew}}, \bibinfo {author} {\bibfnamefont {K.}~\bibnamefont {Burke}}, \
  and\ \bibinfo {author} {\bibfnamefont {M.}~\bibnamefont {Ernzerhof}},\ }\href
  {https://doi.org/10.1103/PhysRevLett.77.3865} {\bibfield  {journal} {\bibinfo
   {journal} {Phys. Rev. Lett.}\ }\textbf {\bibinfo {volume} {77}},\ \bibinfo
  {pages} {3865} (\bibinfo {year} {1996})}\BibitemShut {NoStop}%
\bibitem [{\citenamefont {Hamann}(2013)}]{hamann2013optimized}%
  \BibitemOpen
  \bibfield  {author} {\bibinfo {author} {\bibfnamefont {D.~R.}\ \bibnamefont
  {Hamann}},\ }\href {https://doi.org/10.1103/PhysRevB.88.085117} {\bibfield
  {journal} {\bibinfo  {journal} {Phys. Rev. B}\ }\textbf {\bibinfo {volume}
  {88}},\ \bibinfo {pages} {085117} (\bibinfo {year} {2013})}\BibitemShut
  {NoStop}%
\bibitem [{\citenamefont {Schlipf}\ and\ \citenamefont
  {Gygi}(2015)}]{schlipf2015optimization}%
  \BibitemOpen
  \bibfield  {author} {\bibinfo {author} {\bibfnamefont {M.}~\bibnamefont
  {Schlipf}}\ and\ \bibinfo {author} {\bibfnamefont {F.}~\bibnamefont {Gygi}},\
  }\href {https://doi.org/10.1016/j.cpc.2015.05.011} {\bibfield  {journal}
  {\bibinfo  {journal} {Comput. Phys. Commun.}\ }\textbf {\bibinfo {volume}
  {196}},\ \bibinfo {pages} {36} (\bibinfo {year} {2015})}\BibitemShut
  {NoStop}%
\bibitem [{\citenamefont {Gonze}\ \emph {et~al.}(2020)\citenamefont {Gonze},
  \citenamefont {Amadon}, \citenamefont {Antonius}, \citenamefont {Arnardi},
  \citenamefont {Baguet}, \citenamefont {Beuken}, \citenamefont {Bieder},
  \citenamefont {Bottin}, \citenamefont {Bouchet}, \citenamefont {Bousquet}
  \emph {et~al.}}]{ABINIT}%
  \BibitemOpen
  \bibfield  {author} {\bibinfo {author} {\bibfnamefont {X.}~\bibnamefont
  {Gonze}}, \bibinfo {author} {\bibfnamefont {B.}~\bibnamefont {Amadon}},
  \bibinfo {author} {\bibfnamefont {G.}~\bibnamefont {Antonius}}, \bibinfo
  {author} {\bibfnamefont {F.}~\bibnamefont {Arnardi}}, \bibinfo {author}
  {\bibfnamefont {L.}~\bibnamefont {Baguet}}, \bibinfo {author} {\bibfnamefont
  {J.-M.}\ \bibnamefont {Beuken}}, \bibinfo {author} {\bibfnamefont
  {J.}~\bibnamefont {Bieder}}, \bibinfo {author} {\bibfnamefont
  {F.}~\bibnamefont {Bottin}}, \bibinfo {author} {\bibfnamefont
  {J.}~\bibnamefont {Bouchet}}, \bibinfo {author} {\bibfnamefont
  {E.}~\bibnamefont {Bousquet}},  \emph {et~al.},\ }\href
  {https://doi.org/10.1016/j.cpc.2019.107042} {\bibfield  {journal} {\bibinfo
  {journal} {Comput. Phys. Commun.}\ }\textbf {\bibinfo {volume} {248}},\
  \bibinfo {pages} {107042} (\bibinfo {year} {2020})}\BibitemShut {NoStop}%
\bibitem [{\citenamefont {Novoselov}\ \emph {et~al.}(2005)\citenamefont
  {Novoselov}, \citenamefont {Jiang}, \citenamefont {Schedin}, \citenamefont
  {Booth}, \citenamefont {Khotkevich}, \citenamefont {Morozov},\ and\
  \citenamefont {Geim}}]{Novoselov10451}%
  \BibitemOpen
  \bibfield  {author} {\bibinfo {author} {\bibfnamefont {K.~S.}\ \bibnamefont
  {Novoselov}}, \bibinfo {author} {\bibfnamefont {D.}~\bibnamefont {Jiang}},
  \bibinfo {author} {\bibfnamefont {F.}~\bibnamefont {Schedin}}, \bibinfo
  {author} {\bibfnamefont {T.~J.}\ \bibnamefont {Booth}}, \bibinfo {author}
  {\bibfnamefont {V.~V.}\ \bibnamefont {Khotkevich}}, \bibinfo {author}
  {\bibfnamefont {S.~V.}\ \bibnamefont {Morozov}}, \ and\ \bibinfo {author}
  {\bibfnamefont {A.~K.}\ \bibnamefont {Geim}},\ }\href {\doibase
  10.1073/pnas.0502848102} {\bibfield  {journal} {\bibinfo  {journal} {Proc.
  Natl. Acad. Sci. U.S.A.}\ }\textbf {\bibinfo {volume} {102}},\ \bibinfo
  {pages} {10451} (\bibinfo {year} {2005})}\BibitemShut {NoStop}%
\bibitem [{\citenamefont {Balendhran}\ \emph {et~al.}(2015)\citenamefont
  {Balendhran}, \citenamefont {Walia}, \citenamefont {Nili}, \citenamefont
  {Sriram},\ and\ \citenamefont {Bhaskaran}}]{balendhran2015elemental}%
  \BibitemOpen
  \bibfield  {author} {\bibinfo {author} {\bibfnamefont {S.}~\bibnamefont
  {Balendhran}}, \bibinfo {author} {\bibfnamefont {S.}~\bibnamefont {Walia}},
  \bibinfo {author} {\bibfnamefont {H.}~\bibnamefont {Nili}}, \bibinfo {author}
  {\bibfnamefont {S.}~\bibnamefont {Sriram}}, \ and\ \bibinfo {author}
  {\bibfnamefont {M.}~\bibnamefont {Bhaskaran}},\ }\href
  {https://doi.org/10.1002/smll.201402041} {\bibfield  {journal} {\bibinfo
  {journal} {Small}\ }\textbf {\bibinfo {volume} {11}},\ \bibinfo {pages} {640}
  (\bibinfo {year} {2015})}\BibitemShut {NoStop}%
\bibitem [{\citenamefont {Huang}\ \emph {et~al.}(2006)\citenamefont {Huang},
  \citenamefont {Wu},\ and\ \citenamefont {Hwang}}]{huang2006thickness}%
  \BibitemOpen
  \bibfield  {author} {\bibinfo {author} {\bibfnamefont {Y.}~\bibnamefont
  {Huang}}, \bibinfo {author} {\bibfnamefont {J.}~\bibnamefont {Wu}}, \ and\
  \bibinfo {author} {\bibfnamefont {K.-C.}\ \bibnamefont {Hwang}},\ }\href
  {https://doi.org/10.1103/PhysRevB.74.245413} {\bibfield  {journal} {\bibinfo
  {journal} {Phys. Rev. B}\ }\textbf {\bibinfo {volume} {74}},\ \bibinfo
  {pages} {245413} (\bibinfo {year} {2006})}\BibitemShut {NoStop}%
\bibitem [{Note3()}]{Note3}%
  \BibitemOpen
  \bibinfo {note} {The value is also more than two times that computed from DFT
  using the atomic dipole model \cite {dumitricua2002curvature}, which requires
  an artificial partitioning of the electron density.}\BibitemShut {Stop}%
\bibitem [{\citenamefont {Nikiforov}\ \emph {et~al.}(2014)\citenamefont
  {Nikiforov}, \citenamefont {Dontsova}, \citenamefont {James},\ and\
  \citenamefont {Dumitric{\u{a}}}}]{nikiforov2014tight}%
  \BibitemOpen
  \bibfield  {author} {\bibinfo {author} {\bibfnamefont {I.}~\bibnamefont
  {Nikiforov}}, \bibinfo {author} {\bibfnamefont {E.}~\bibnamefont {Dontsova}},
  \bibinfo {author} {\bibfnamefont {R.~D.}\ \bibnamefont {James}}, \ and\
  \bibinfo {author} {\bibfnamefont {T.}~\bibnamefont {Dumitric{\u{a}}}},\
  }\href {https://doi.org/10.1103/PhysRevB.89.155437} {\bibfield  {journal}
  {\bibinfo  {journal} {Phys. Rev. B}\ }\textbf {\bibinfo {volume} {89}},\
  \bibinfo {pages} {155437} (\bibinfo {year} {2014})}\BibitemShut {NoStop}%
\bibitem [{\citenamefont {Kundalwal}\ \emph {et~al.}(2017)\citenamefont
  {Kundalwal}, \citenamefont {Meguid},\ and\ \citenamefont
  {Weng}}]{kundalwal2017strain}%
  \BibitemOpen
  \bibfield  {author} {\bibinfo {author} {\bibfnamefont {S.}~\bibnamefont
  {Kundalwal}}, \bibinfo {author} {\bibfnamefont {S.}~\bibnamefont {Meguid}}, \
  and\ \bibinfo {author} {\bibfnamefont {G.}~\bibnamefont {Weng}},\ }\href
  {https://doi.org/10.1016/j.carbon.2017.03.013} {\bibfield  {journal}
  {\bibinfo  {journal} {Carbon}\ }\textbf {\bibinfo {volume} {117}},\ \bibinfo
  {pages} {462} (\bibinfo {year} {2017})}\BibitemShut {NoStop}%
\end{thebibliography}
%

\end{document}